\pdfoutput=1

\documentclass[reprint,aps,prl,showpacs,amsfonts,twocolumn]{revtex4-1}
\usepackage{newtxtext}
\usepackage{epsfig,amsopn}
\usepackage{graphicx}
\usepackage{amsmath,amssymb}
\usepackage{natbib,color,xcolor}
\usepackage{braket}
\usepackage{float}
\usepackage{enumerate}
\usepackage{hyperref}
\allowdisplaybreaks[1]

\newcommand{\eref}[1]{Eq.~(\ref{#1})}%
\newcommand{\fref}[1]{Fig.~\ref{#1}} %

\def\bea{\begin{eqnarray}}
\def\eea{\end{eqnarray}}

\def\Pe{\text{Pe}}
\begin{document}

\title{Landau theory of restart transitions}

\author{{\normalsize{}Arnab Pal$^{1,2,3}$}
{\normalsize{}}}
\email{arnabpal@mail.tau.ac.il}

\author{{\normalsize{}V. V. Prasad$^{4}$}
{\normalsize{}}}
\email{prasad.vv@weizmann.ac.il}

\affiliation{\noindent \textit{$^{1}$School of Chemistry, Raymond and Beverly Sackler Faculty of Exact Sciences, Tel Aviv University, Tel Aviv 6997801, Israel}}

\affiliation{\noindent \textit{$^{2}$Center for the Physics and Chemistry of Living Systems. Tel Aviv University, 6997801, Tel Aviv, Israel}}

\affiliation{\noindent \textit{$^{3}$The Sackler Center for Computational Molecular and Materials Science, Tel Aviv University, 6997801, Tel Aviv, Israel}}

\affiliation{\noindent \textit{$^{4}$Department of Physics of Complex Systems, Weizmann Institute of Science, Rehovot 7610001, Israel}}

\date{\today}

\begin{abstract}
We develop a Landau like theory to characterize the phase transitions in resetting systems. Restart can either accelerate or hinder the completion of a first passage process. The transition between these two phases is characterized by the behavioral change in the order parameter of the system namely the optimal restart rate. Like in the original theory of Landau, the optimal restart rate can undergo a first or second order transition depending on the details of the system. Nonetheless, there exists no unified framework which can capture the onset of such novel phenomena. We unravel this in a comprehensive manner and show how the transition can be understood by analyzing the first passage time moments. Power of our approach is demonstrated in two canonical paradigm setup namely the Michaelis Menten chemical reaction and diffusion under restart.

\end{abstract}


\maketitle

\textit{Introduction}.---
Consider a perpetual non-equilibrium process following any 
generic governing law of motion which gets stopped 
intermittently and reinstated to a pre-selected 
configuration. The process resumes and continues until
the next resetting epoch. This is known as the resetting or 
restart phenomena  \cite{RestartFP1,RestartFP2,RestartFP3,RestartFP4,RestartFP5,RestartFP6,
RestartFP7,RestartFP8,RestartFP9,RestartFP10,RestartFP11,RestartFP12,RestartFP13,RestartFP14,
RestartFP15,RestartFP16,RestartFP17,RestartFP18,RestartFP19,RestartFP20,RestartFP21,RestartFP22,
RestartFP23,RestartFP24,RestartFP25,RestartFP26,RestartFP27,RestartFP28,restart_thermo2}. A fingerprint aspect of the topic that
has gained a lot of attention is its ramifications to any arbitrary first
passage time process \cite{RednerBook,MetzlerBook,Schehr-review,Benichou-review}. Indeed an underlying first passage process
that becomes subject to a constant restart rate can generically show
different behavior since restart can hinder or accelerate
the underlying process. In the former case, optimal restart
rate (ORR) that minimizes the mean first passage time (MFPT) is
trivially fixed at zero while in the latter case one can always obtain a non-zero restart rate which
will minimize the MFPT and thus the ORR is
finite. This means that by tuning the restart rate to this
optimal value, the completion can be expedited. No surprise that
this crucial observation has taken the center stage and lead to a myriad of studies \cite{RestartFP1,RestartFP2,RestartFP18,RestartFP19,RestartFP20,RestartFP21,RestartFP22,
RestartFP23,RestartFP24,RestartFP25,RestartFP26,RestartFP27,RestartFP28}.

Although most of the current investigations focused on whether
restart overperforms or underperforms the completion, 
sharp identification of the behavioral transition that
accompanies this is quite less understood.
We argue that this transition is in the 
same genre of canonical phase transition prevalent in equilibrium 
statistical physics and the ORR naturally roles the play of the order parameter. To be specific, a finite value of the ORR indicates that restart can be speed-up the completion while a zero value of the same simply indicates that restarting the process will leave unchanged or hinder the completion. Restart transitions can be both of first and second order transition like in the classical phase transition. In the former case, the ORR will have a discontinuous jump while in the latter case, it continuously vanishes to zero. First order transition in the ORR rate was observed in the case of a non-diffusive search with the aid of L\'evy flight \cite{RestartFP17} or a search comprising of a drifted random walker with exponential flights subjected to restart \cite{RestartFP21}. The continuous transition in the restart rate, on the other hand, was observed in a system of Brownian walker in the presence of potential subjected to resetting \cite{RestartFP25,RestartFP26}. Nonetheless there is no generic understanding of how the ORR behaves close to the transition with respect to the system parameters and what precisely characterizes the \textit{nature} of the transition. To understand and  characterize this non-trivial behavior, we develop a Landau like theory and provide a comprehensive picture of generic restart transitions.

The phenomenological theory developed by Landau describes the universal behavior of a system near the phase transition \cite{PT1,PT2,PT3}. According to this theory one can write down the free-energy of a system as a polynomial in the order parameters with the coefficients which appropriately describe the phase transition. In one phase, the order parameter is finite while in the other it vanishes e.g., magnetization in the classical ferromagnetic system. Following Landau theory, we express the MFPT as a power series in terms of the restart rate $r$ near the transition, and show how by utilizing the relations between the coefficients, it is possible to predict the emergence of first and second order transition. 
We start by writing the MFPT under restart as a polynomial in restart rate
\bea
\mathcal{T}(r)=a_0+a_1 r+a_2 r^2+a_3 r^3+ \cdots~,
\label{L1}
\eea
where $a_i$-s are the expansion coefficients and will play the key role to determine the transitions.
Such kind of expansion respects a set of postulates which we will state now. By taking $r$ strictly to be zero, we see that $\mathcal{T}=a_0$ which is the MFPT without restart and is assumed to be finite. The expansion also assumes that $\mathcal{T}$ is an analytic function of both $r$ and the coefficients. 
Since both $\mathcal{T}$ and $r$ are strictly non-negative quantities, the system does not have any symmetry around $r=0$, and thus can have all order terms in $r$. Importantly, the coefficient of the highest order in $r$ must also be positive, otherwise one can minimize $\mathcal{T}$ by $r \to \infty$. We will now extensively explore how using the expansion as stated above one can capture the generic features of first and second order transitions.

\textit{Second order phase transition}.---
The second order phase transition is characterized by the continuous vanishing of the order parameter near the critical point. When the ORR
goes continuously to zero there exists a range close to the transition point where only the first few terms of the power series are relevant. 
 Here we keep only till the third order term in the polynomial and show that
 the second order phase transition is determined by the slope of the MFPT near $r\to 0$ i.e., by the change of sign $a_1$ in \eref{L1} where $a_0 \geq 0$ and $a_2>0, a_3 >0$. To show this, we first obtain the ORR by setting
\bea
\frac{\partial \mathcal{T}}{\partial r}\bigg|_{r=r_s}=0~,
\label{second-min}
\eea
where $r_s$ is the optimal restart rate.
Using \eref{L1} and \eref{second-min}, we find $a_1+2a_2r_s+3a_3r_s^{2}=0$. The valid solution of the equation, when $a_1<0$, is given by 
$r_s= \frac{a_2}{3a_3}\left[\sqrt{1+3|a_1|a_3/a_2^2}-1  \right]>0$. Thus, there exists a finite restart rate which can minimize the MFPT. As we tune the parameter to be $a_1\geq 0$, the ORR is shifted towards $r_s=0$ which means restart will not quicken the completion. Close to the critical point, the approach of ORR continuously to the
value zero is captured by doing an expansion around $a_1=0$ which gives $r_s \sim |a_1|/2a_2$. To summarize, the continuous transition is characterized by
\begin{equation}
\begin{array}{l}
r_s=\left\{ \begin{array}{lll}
0 &  & \text{if ~~}a_1 \geq 0\text{ }\\
 & \text{ \ \ }\\
|a_1|/2a_2&  & \text{if~~ }a_1<0\text{ ,}
\end{array}\right.\text{ }\end{array}
\label{second-order-transition}
\end{equation}
Since $r_s$ behaves linearly near the phase transition, in the the ordered phase the ORR will have a universal power law dependency 
\bea
r_s \propto |a_1|^{\beta}~,~~~\text{where}~~~~\beta=1~.
\label{critical-exponent}
\eea
This relation is reminiscent of the same observed in classical systems e.g., liquid-gas or ferromagnetic systems where $\beta$ is found to be universal.

\textit{First order phase transition}.---
Landau's theory can also be useful to explain the first order transition provided the expansion is justified and this is so when the order parameter is arbitrarily small. To understand this transition in our context, let's again consider the MFPT only till the third order term in the series such that $\mathcal{T}(r)=a_0+a_1 r+a_2 r^2+a_3 r^3$, where $a_1>0$ and $a_3>0$. Given the expansion, the sign of $a_2$ decides whether there will be a first order transition. If $a_2 \geq 0$, the ORR is fixed at zero since in this case $\mathcal{T}$ monotonically increases as a function of restart rate. On the other hand when $a_2<0$, it is possible to have a finite ORR. The first order transition is then characterized by two simple relations
\bea
\mathcal{T}(r_f)=\mathcal{T}(0)~,~~~~~~~
\frac{\partial \mathcal{T}}{\partial r}\bigg|_{r=r_f}=0~,
\label{rf}
\eea
where $r_f$ is the optimal restart rate. From \eref{rf}, we find
$a_1r_f+a_2 r_f^{2}+a_3 r_f^{3}=0 $ and
$a_1+2a_2 r_f+3a_3 r_f^{2}=0$.
Solving these two equations we obtain
\bea
r_f=-\frac{a_2}{2a_3},~~~~~a_1=\frac{a_2^2}{4a_3}
\label{r_f_a_1}
\eea
Unlike the continuous transition, first order transition is characterized by a finite $a_1$ and at this transition point the ORR experiences a discontinuity. The locus of the ORR as a function of $a_1$ can be obtained by solving \eref{rf}. This gives $r_f=\frac{|a_2|}{3a_3}\left[1+\sqrt{1-3a_1a_3/a_2^2}  \right]>0$ which at the transition point $a_1=a_2^2/4a_3$ terminates at a value $|a_2|/2a_3$ and jumps discontinuously to zero. Furthermore, the discontinuity in ORR at the transition point can be captured in a single equation
\begin{equation}
\begin{array}{l}
r_f=\left\{ \begin{array}{lll}
0 &  & \text{if ~~}a_2 \geq 0\text{ }\\
 & \text{ \ \ }\\
|a_2|/2a_3&  & \text{if~~ }a_2<0\text{ ,}
\end{array}\right.\text{ }\end{array}
\label{first-order-transition}
\end{equation}
where the ORR can be seen to follow a similar power law dependency $r_f \propto |a_2|^{\gamma}$, with $\gamma=1$.

Treating first order transitions using Landau expansion reveals a significant caveat. If the first order transition involves a discontinuous jump from zero to a finite value significantly
away from zero, it cannot be captured by a small order expansion of the MFPT near $r = 0$. In fact, in this case, one also has to consider the possibility that higher order terms should be included in the power series expansion depending on the problem. It is clear, however, that this expansion becomes more
respectable as the discontinuity in the order parameter becomes smaller,
particularly when it goes to zero. Thus the theory is expected to
describe well when 
\bea
a_1=0,~~~~~a_2=0~,
\eea
which is the tricritical point, where the changeover between the first and second order phase transition takes place.

\begin{figure*}[t]
\includegraphics[height=4.5cm,width=17.5cm]{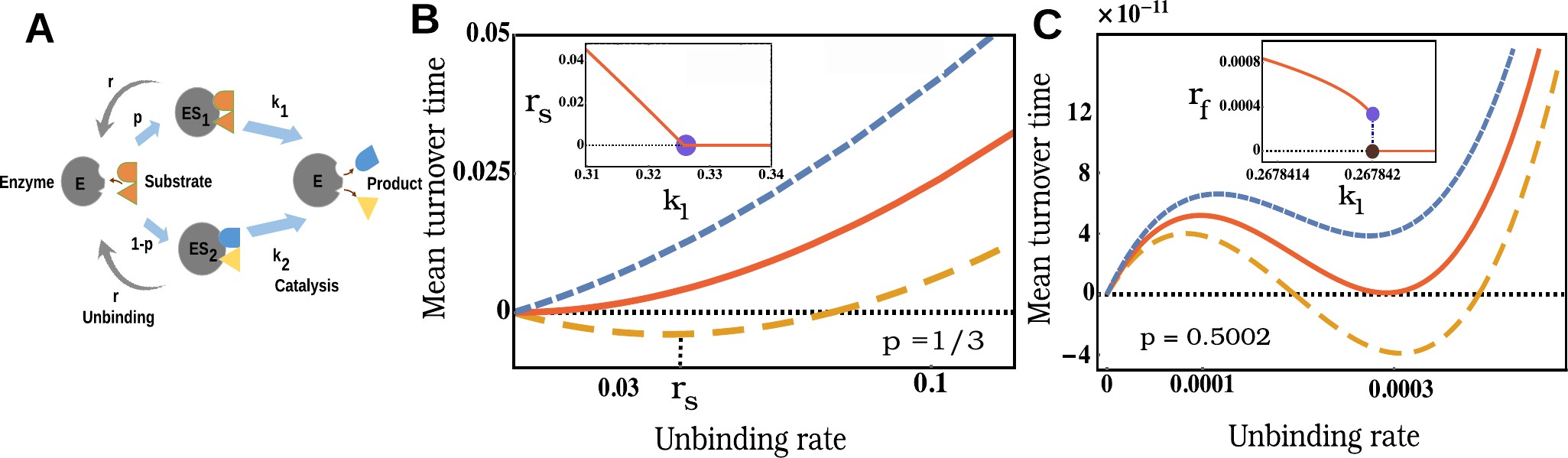}
\caption{Michalis-Menten reaction setup and the transitions.
Panel A illustrates the basic steps of the Michaelis-Menten chemical reaction. Panel B plots the mean turnover time as a function of unbinding rate $r$ for $k_2=1,p=1/3$ for different values of $k_1$. The minima $r_s$ (shown by the dashed vertical line) corresponds to the ORR which continuously vanishes to zero at the critical point $k_1=0.3258$ (in blue marker) as shown in the inset. Panel C plots the mean turnover time as a function of unbinding rate $r$ when $k_2=1,p=0.5002$ while varying $k_1$. The ORR $y_f$ exhibits a discontinuous transition at $k_1=0.267842$ (in solid marker) as shown in the inset.
}
\label{Fig1}
\end{figure*}

\textit{Physical meaning of the coefficients}.--- So far our analysis has been restricted to the generic coefficients $a_i$-s without providing information how they are related to the system parameters. To bridge this gap, we first note that MFPT for any generic first passage time (denoted by $T$) under constant restart rate $r$ is given by $\mathcal{T}=\frac{1-\tilde{T}(r)}{r\tilde{T}(r)}$ \cite{RestartFP22,RestartFP23},
where $\tilde{T}(r)=\langle e^{-rT} \rangle$ is the Laplace transformation of the underlying first passage time $T$ evaluated at $r$. A simple Taylor's series expansion of the above expression around $r=0$ then connects the coefficients $a_i$-s to the moments of the underlying process. In particular, comparing the Taylor's series expansion and \eref{L1}, we identify $a_0=\langle T \rangle,~a_1= -\frac{\langle T^2 \rangle}{2} + \langle T \rangle^2,~a_2= \frac{1}{6} \langle T^3 \rangle +\langle T \rangle^3-\langle T \rangle\langle T^2 \rangle,~a_3=-\frac{\langle T^4 \rangle}{4!} +\frac{\langle T^3 \rangle\langle T \rangle}{3}+\frac{\langle T^2 \rangle^2}{4}-\frac{3\langle T^2 \rangle\langle T \rangle^2}{2}+\langle T \rangle^4$, and so on \cite{SM}. The moments (hence the coefficients) are explicit functions of the system parameters, and thus the transitions can be explicitly characterized in terms of them. Recall that the second order phase transition is characterized by the change in sign of $a_1$, which, in terms of the moments would imply whether the coefficient of variation, which stands for the ratio between the standard deviation and the mean of $T$, is higher or lower than unity \cite{RestartFP22,RestartFP23,RestartFP27}. We find that this is only a sufficient condition for the second order \textit{albeit} the first order where the change in sign of $a_2$ plays the pivotal role. To illustrate how the frame-work developed above can be realized in practice, we now investigate two case studies namely the Michaelis-Menten reaction scheme, and diffusion under restart.

\textit{Michaelis-Menten chemical reaction}.---
Bio-chemical processes such as Michaelis-Menten kind reactions are an integral part of first passage under restart formalism \cite{Restart-Biophysics1,Restart-Biophysics2,Restart-Biophysics3,Restart-Biophysics4,Restart-Biophysics5}. 
According to this reaction scheme, an enzyme (E) binds with a substrate (S) to form a metastable complex (ES). However, this complex can be formed in two alternative states ES$_1$ with probability $p$ or ES$_2$ otherwise. From here, two things can happen: either of the complexes can be converted into a product (P) via catalysis which marks the turnover of an enzymatic reaction cycle (see \fref{Fig1}A) or the enzyme unbinds. Thus the reaction will resume again. The catalytic process differ between the two states through their catalytic rates which are correspondingly given by $k_1$ and $k_2$. Here we assume the catalytic time distribution to be $f_{T_{cat}}(t)=p k_1^2 te^{-k_1t}+(1-p)~k_2^2~te^{-k_2t}$, which is also known as the hyper-Erlang distribution \cite{Restart-Biophysics1,Restart-Biophysics2,Restart-Biophysics3}.
However, the states ES$_1$ or ES$_2$ have the same substrate unbinding rate $r$ so that the unbinding time distribution is taken from an exponential distribution $f_{T_{off}}(t)=re^{-rt}$. 
Here, we also assume a steady state concentration of the substrate in the solution so that the time spent to form the metastable states can safely be neglected. Given the distributions,
one can immediately use the tools from the theory of first passage under restart to obtain the mean turnover time which gives \cite{SM}
\bea
\mathcal{T}=\frac{(k_1+r)^2\left[(k_2+r)^2-(1-p)k_2^2\right]-pk_1^2(k_2+r)^2 }{r 
\left[ pk_1^2(k_2+r)^2+(1-p)k_2^2(k_1+r)^2 \right]}\hspace{0.75cm}
\label{MTT}
\eea
Expanding $\mathcal{T}$ around $r=0$ as in \eref{L1}, one can immediately identify $a_0=2\left[k_1+(k_2-k_1)p \right]/k_1k_2,~a_1=[3k_1^2(-1+p)-3k_2^2p+4(k_1+(k_2-k_1)p)^2]/(k_1 k_2)^2$,
and others we reserve in \cite{SM}. The second order transition is determined by $a_1=0$ which implies $3k_1^2(-1+p)-3k_2^2p+4(k_1+(k_2-k_1)p)^2=0$. For fixed $k_2$ and $p$, the above equation gives a solution for $k_1$. To see the transition, we first plot the mean turnover time (shifted by $a_0$ which is the underlying MFPT) as a function of the unbinding rate $r$ while varying the parameter $k_1$ for $k_2=1,p=1/3$ in \fref{Fig1}B. As we vary $k_1$, the ORR $r_s$ continuously vanishes to zero from a finite value. The transition occurs exactly at $k_1=0.3258$ (indicated by marker in \fref{Fig1}B inset), which can also be found by solving the above equation by substituting $k_2=1,p=1/3$. Importantly, $r_s$ vanishes linearly to zero as a function of $k_1$ near the transition point as predicted from \eref{critical-exponent}. Indeed this is a signature of second order transition and such scaling was also observed phenomenologically in other set-up \cite{RestartFP25,RestartFP26}.

\begin{figure}[b]
\includegraphics[height=5cm,width=7.05cm]{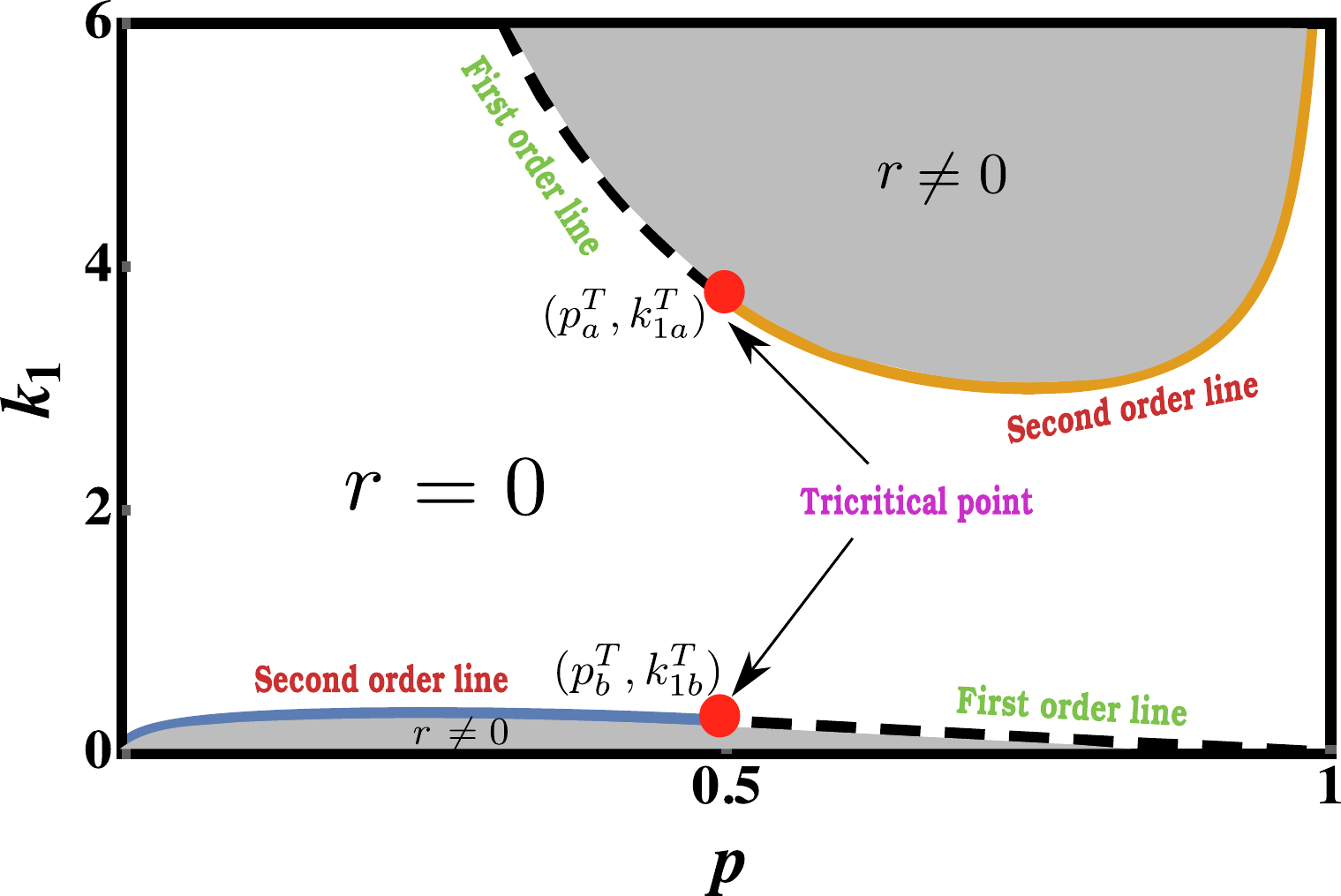}
\caption{The phase diagram for the Michelis-Menten reaction scheme.
The plot shows the phase diagram in the ($p,k_1$) plane for a fixed $k_2=1$.
The grey  shaded region indicate the region in the parameter space for which
ORR is non-zero. In the non-shaded region ORR is zero. First order transition lines between the two regions is indicated by black dashed line. The solid lines corresponds to the continuous transition lines. The tricritical points where the two lines merge is marked by the red circles.}
\label{erlang-phase}
\end{figure}

\begin{figure}[t]
\includegraphics[height=3.cm,width=8.55cm]{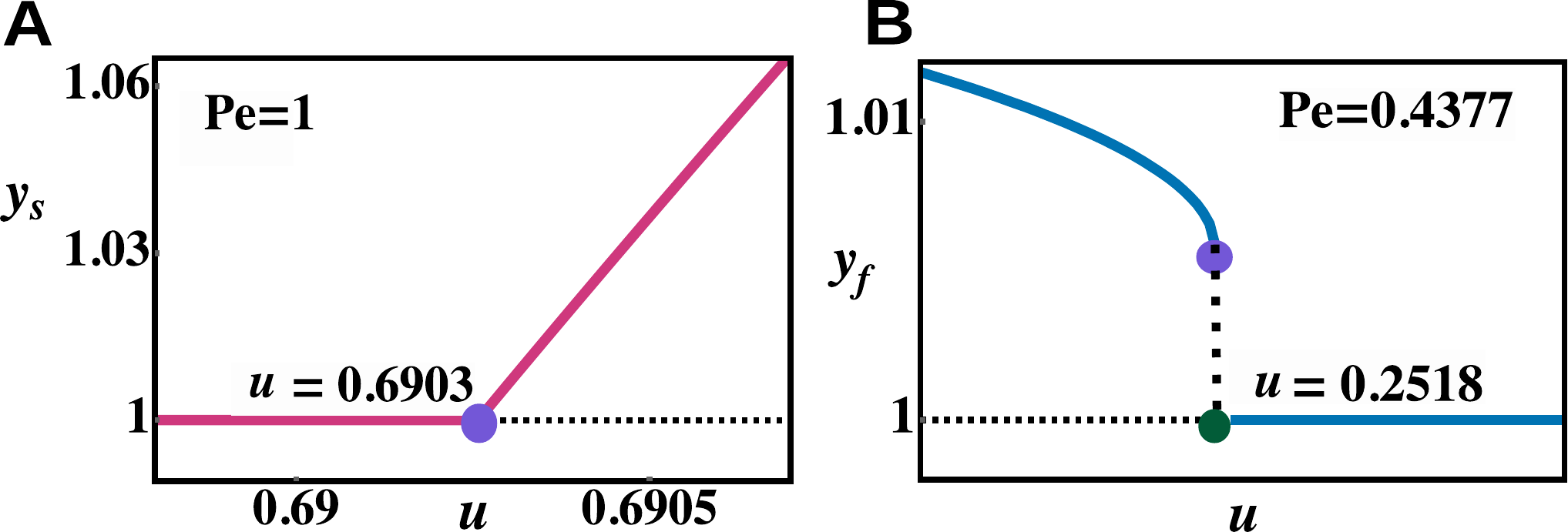}
\caption{Phase transitions in ORR in drift-diffusion under confinement. Panel A: continuous transition of ORR $y_s$ as a function of $u$ for a fixed $\Pe=1.0$. The transition occurs at the critical value $u=0.6903$ (in blue circle). Panel B: first order phase transition in ORR $y_f$ as a function of $u$ for a fixed $\Pe=0.4377$. The ORR jumps from a finite value $1.0086$ to unity at the transition point $u=0.2518$ (in solid marker).}
\label{diffusion-y}
\end{figure}

We now turn our attention to \fref{Fig1}C where we have plotted the mean turnover rate as a function of unbinding rate for different $k_1$ fixing $k_2=1.0$ and $p=0.5002$. As we vary $k_1$, we observe that the ORR jumps from a finite value $r_f=3\times 10^{-4}$ to zero and this transition occurs at $k_1=0.267842$. Note that these values (indicated by markers in \fref{Fig1}C inset) match exactly with those determined from \eref{r_f_a_1} for given parameters thus validating our theory. However, choice of these co-ordinates is not merely coincidental. In fact, these points are chosen at the proximity of the tricritical point
which can be determined by setting $a_1=0,a_2=0$. The second order line in $(p,k_1)$ plane can be determined by setting $a_1=0$ for each fixed $k_2$. This is shown in solid lines in \fref{erlang-phase}. In contrast, to generate the entire first order contour line, one needs to consider the exact mean turnover time (\eref{MTT}) and apply the conditions given by \eref{rf} \cite{SM}. We show these lines by dashed curves in \fref{erlang-phase}.
The first and second order lines meet at the tricritical points which are given by $p=\frac{1}{2},~k_1=k_2(2 \pm \sqrt{3})$ \cite{SM}. Therefore, for each $k_2$, there are two tricritical points in $(p,k_1)$ plane which are denoted by $(p_a^T,k_{1a}^T)$ and $(p_b^T,k_{1b}^T)$ and in \fref{Fig1}C, we have chosen our co-ordinates near the second critical point. The phase diagram comprising first and second order phase transitions in the entire parameter space is illustrated in \fref{erlang-phase}.

\begin{figure}[b]
\includegraphics[height=6.5cm,width=8.55cm]{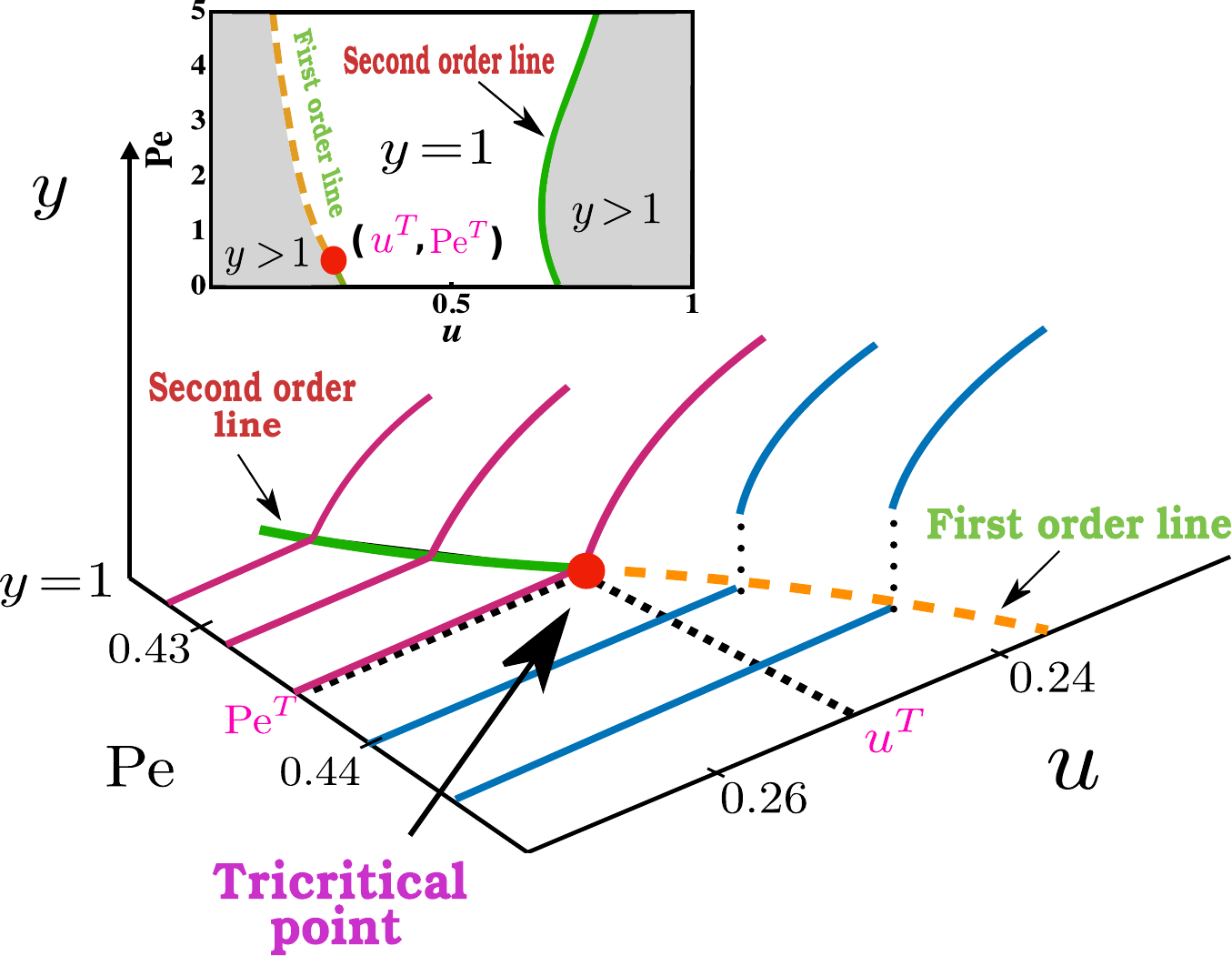}
\caption{Phase transitions in ORR in a drift-diffusion process. The ORR is plotted as a function of $\Pe$ and $u$  close to the tricritical point (marked in red circle). The first and second order transition lines are marked by the orange (dashed) line and the solid green line respectively and they meet at the tricritical point denoted by $u^T=0.25186,~\Pe^T=0.43764$.
 An extended phase diagram of the same setup for the entire range of $u$ is depicted in the inset. The shaded regions represent those with $y>1$, and the non-shaded
one with $y=1$. At large $\Pe$, the continuous transitions occur close to $u=1$ while the first order transition is seen strictly near $u=0$.}
\label{diffusion-phase}
\end{figure}

\textit{Diffusion under restart}.---
We consider motion of a drifted Brownian particle (diffusion constant $D$) confined in a one-dimensional domain with absorbing boundaries at $[0,L]$. Starting from $x_0$ at time zero, the particle is drifted towards the right boundary $L$ with velocity $v>0$. In addition, the particle is also reset back to $x_0$ with a constant rate $r$. The mean first passage time for the particle to escape any of the boundaries is then found to be \cite{SM}
\bea
\mathcal{T}=\frac{L^2}{D \Pe^2}~ \mathcal{G}(y,u,\Pe)
\eea
where we have defined $\Pe=\frac{Lv}{2D}$, which is the P\'eclet number and $ y=\sqrt{1+4Dr/v^2},~u=\frac{x_0}{L}$. The scaling function $\mathcal{G}$ is given by
{\footnotesize\bea
\mathcal{G}(y,u,\Pe)=\frac{1}{y^2-1} \left( \frac{e^{u \Pe} \sinh \left[y \Pe \right]}{\sinh \left[ (1-u)y \Pe\right]+e^{\Pe}\sinh \left[ uy \Pe \right]}-1\right) \hspace{0.5cm}\eea}
To get the ORR, we do the following optimization
$\partial_y \mathcal{G}(y,u,\Pe)=0$.
Since $y$ is a measure of the ORR here, $r \to 0$ would automatically imply the limit $y \to 1$. Expanding $\mathcal{G}$ around $y=1$ immediately reveals the coefficients which we present in \cite{SM}. Setting $a_1=0$ we immediately obtain a relation between $\Pe$ and $u$, which in turn determines the critical point where the second order transition occurs. To see this transition, we have plotted the ORR as a function of $u$ while fixing $\Pe$ in \fref{diffusion-y}A. This shows that $y_s$ continuously vanishes zero exactly at $u=0.6903$ (indicated by blue marker in \fref{diffusion-y}A) as predicted from the theory. Note that if $u>0.6903$, we will always find $y_s>1$ (i.e., restart will be beneficial); otherwise it will be detrimental.
Interestingly, it is also possible to get a simple expression for this critical $u$ in certain limits. To extract this let us take the limits $\Pe \gg 1$ and $u \to 1$ in the expansion of $\mathcal{G}$ which gives
$\mathcal{G} \sim \frac{1}{y^2-1} \left[ e^{(u+y-1-uy)\Pe}-1 \right]$.
Expanding this around $y \to 1$ yields
$ \mathcal{G} \sim \frac{1}{2}\Pe(1-u)+\frac{1}{4}\left[ (\Pe-u \Pe)^2-\Pe(1-u) \right](y-1)+\cdots$, where
the first term on the RHS gives the underlying MFPT given by $ \frac{L^2}{2D\Pe}(1-u)$ \cite{RednerBook}.
Setting the criterion for the second order transition namely $a_1<0$, we obtain $(\Pe-u \Pe)^2-\Pe(1-u) <0$ which gives the following relation $u>u^*=1-\frac{1}{\Pe}$. Hence, for large $\Pe \gg 1$, there is always a finite ORR if $u>u^*$; otherwise it is zero and the critical point is located at $u^*=1-\frac{1}{\Pe}$. As predicted from the theory, the ORR $y_s$ scales linearly with $u$ close to the critical point. This can also be seen from \fref{diffusion-y}A.

The set up also exhibits a rich first order transition phenomena. To capture this transition, 
we once again examine the coordinates near the tricritical point $u^T=0.25186,\Pe^T=0.43764$ (marked by red circle in \fref{diffusion-phase}) which can be computed by setting $a_1=0,a_2=0$. In \fref{diffusion-phase}, we have plotted the ORR as a function of $u$ and $\Pe$. We see that for $u<u^T$ and $\Pe>\Pe^T$, the system undergoes a series of first order transitions 
while the complementary limits show transitions of second order. We choose the coordinates $u=0.25186$ and $\Pe=0.4377$ at the proximity of the tricritical point where the theory provides an excellent match. At these points, the ORR $y_f$ jumps from the value $1.0086$ (marked in blue circle in \fref{diffusion-y}B) to unity discontinuously. This is demonstrated  in \fref{diffusion-y}B. The value of the ORR is in commensurate with the theoretical predictions given by \eref{rf}. Although it is not possible to understand the universal behavior of the first order transitions away from the tricritical point within our framework, one can still capture the entire phase diagram in the $(\Pe,u)$ parameter space by performing the same analysis as was done in the case of Micahelis-Menten reaction scheme. The phase space is demonstrated in the inset of \fref{diffusion-phase}.

\textit{Discussion and outlook}.---
Landau's theory of free energy expansion is a benchmark paradigm 
in equilibrium statistical physics. In this paper, we have unveiled a striking connection between this and the theory of restart transition. We show that the former serves as a natural setting to study the phase transitions in optimal restart rate and crossover between the phases in the parameter space can entirely be characterized by the underlying first passage process. We observe that the well known criterion in terms of the coefficient of variation for optimality is a sufficient condition for the continuous phase transition, but not for the first order. In fact, the discontinuous transitions can take place even if this criterion is not satisfied, and this we find to be a hallmark feature of this transition. The unified framework developed herein allows one to make these generic statements as well as capture the fundamental aspects of the phase transition phenomena in restarted processes \`a la Landau. 

\textit{Contributions}.--- Arnab Pal and V. V. Prasad have equally contributed to this work.

\textit{Acknowledgements}.---
Arnab Pal gratefully acknowledges support from the Raymond and Beverly Sackler Post-Doctoral Scholarship.

\end{document}